\documentclass[a4paper,11pt]{article}

\usepackage[tmargin=1in,bmargin=1in,lmargin=1.in,rmargin=1.in]{geometry}

\usepackage{graphicx, subfigure}
\usepackage{multirow}
\usepackage{url}
\usepackage{array}
\usepackage{slashbox}
\usepackage{rotating}
\usepackage{amsmath}

\begin{document}
\title{Statistical Review of UK Residential Sector Electrical Loads}
\author{G. Tsagarakis, A. J. Collin and A. E. Kiprakis$^{*}$\\Institute for Energy Systems\\School of Engineering\\The University of Edinburgh\\$^{*}$Corresponding author: Aristides.Kiprakis@ed.ac.uk}
\date{\vspace{-5ex}}
\maketitle
\section{Introduction}
Technological developments and increasing energy efficiency have resulted in significant changes in the power consumption of electrical appliances commonly found in the residential load sector. These changes will alter both the magnitude of power consumption and how the power is consumed, which is only partially captured by available load use statistics. In existing literature, power demand is generally represented by annual consumption statistics which are divided into broad load types, defined by the end-use. This classification normally includes: ''wet'' loads, ''cold'' loads, consumer electronics, information and communication technology (ICT) loads, cooking loads and electrical space/water heating loads. Although these statistics clearly have their merits, they do not contain detailed information on the different appliances within each load type.

To fill this gap, this paper presents a comprehensive statistical review of data obtained from a wide range of literature on the most widely used appliances in the UK residential load sector. This focusses on the individual appliances and begins by consideration of the electrical operations performed by the load. This approach allows for the loads to be categorised based on the electrical characteristics, and also provides information on the reactive power characteristics of the load, which is often neglected from standard consumption statistics. This data is particularly important for power system analysis. In addition to this, device ownership statistics and probability distribution functions of power demand are presented for the main residential loads. Although the data presented is primarily intended as a resource for the development of load profiles for power system analysis, it contains a large volume of information which provides a useful database for the wider research community.

\section{Load categorisation}

\subsection{Power consumption}
One of the key properties of any load is the rated power, as this defines the demand in the power system. However, the power consumption may vary between two devices of similar use, e.g. between two LCD TVs from different manufacturers, and a large number of data sources are required to fully characterise the range of power demand. In this paper, a review of manufactururs' data is processed to present probability distribution functions ($pdf$) for the main UK residential load sectors appliances. In addition to this, the reactive power characteristics are included by consideration of the electrical characteristics of the load.

\subsection{Electrical load model} 
The power demand of the electrical appliances is, generally, a function of the supply conditions. This is normally given as a function of the supply voltage magnitude, which may vary in the range of plus minus 10\% from nominal. Traditionally, load characteristics are defined as one of three general load models:

\begin{itemize}
\item Constant impedance - for which the active and reactive power demands of the load vary proportionally to the square of the supply voltage magnitude.
\item Constant current load - for which the active and reactive power demands of the load vary in direct proportion (i.e. linearly) to the supply voltage magnitude.
\item Constant power load - for which the load will draw constant active and reactive power, irrespective of the changes in supply voltage magnitude. 
\end{itemize}

Modern, non-linear loads are better represented as a mixture of these load models, and the relationship can be described by the polynomial load model, (\ref{polmodelact}) for active power and (\ref{polmodelreact}) for reactive power as function of the supply voltage.

\begin{equation}
\label{polmodelact}
P = P_{0}\left[Z_{p}\left(\frac{V}{V_{p}}\right)^{2} + I_{p}\left(\frac{V}{V_{0}}\right) + P_{p}\right]
\end{equation}

\begin{equation}
\label{polmodelreact}
Q = Q_{0}\left[Z_{q}\left(\frac{V}{V_{0}}\right)^{2} + I_{q}\left(\frac{V}{V_{0}}\right) + P_{q}\right]
\end{equation}

\noindent where: $P$, $Q$ and $V$ are the actual active and reactive power demand and supply system voltage accordingly while $P_{0}$, $Q_{0}$ and $V_{0}$ are the nominal/rated corresponding values. $Z_{p}$, $I_{p}$, $P_{p}$ and $Z_{q}$, $I_{q}$, $P_{q}$ are the polynomial load model coefficients.\\

In this review, the electrical characteristics are included by considering the actual electrical processes performed in each load. This allows the conversion of the load types into load categories, which better represent the electrical behaviour of the loads. The six main load types (''wet'', ''cold'', consumer electronics (CE), ICT, cooking and space/water heating) can be converted into five general load categories:

\renewcommand{\labelenumi}{\alph{enumi})}
\begin {enumerate}
\item Lighting loads with three subcategories: general incandescent lamps (GIL) or lamps with similar electrical behaviour, i.e. halogen lamps (HIL), energy efficient lighting: compact and linear fluorescent lamps (CFL/LFL) and light-emitting diode (LED) light sources (LED LS).
\item Resistive loads: appliances that are assumed to behave as ideal resistance. Examples of this category include electrical heating units and electrical cookers.
\item Power electronics loads: including all consumer electronics and information and communication technology (ICT) appliances. These loads are sensitive to voltage variations and require a regulared dc voltage suppy, commonly referred to as a switch-mode power supply (SMPS). These loads will draw non-linear current waveforms from the supply systema and may require additional power factor correction (PFC) circuits to satisfy harmonic legislation, given in \cite{PFCleg}. This effectively creates three sub categories of SMPS loads: a) loads of rated power less than 75 W which do not need any power factor correction (PFC), b) loads with passive PFC (p-PFC), c) and loads with active PFC (a-PFC).
\item Directly-connected motors: which are utilised in a large number of domestic appliances, particulary in ''wet'' and ''cold'' loads. In the UK residential sector, these are predominantly single-phase induction motors (SPIM). There are possible sub-categories based on the electrical configuration of the SPIM and the type of mechanical loading on the motor. In the UK residential sector these possible variations can be classified as: the addition of run capacitor (capacitor start-run - CSR) or not (resistive start-inductor run - RSIR) and constant or quadratic torque (CT and QT, respectively).
\item Drive-controlled motors: are more efficient and allow for a better control of the torque/speed performance. Although they are more prevalent in other load sectors, they are not widely found in the UK residential load sector. However, they are expected to increase in numbers in the next generation of energy efficient domestic appliances.
\end {enumerate}

Further information on the load categories classification can be found in \cite{collin,cresswell}. A summary of the electrical load models of the load categories is given in Table \ref{models_table}.

\begin {table}[h]
\centering
\caption {Models of load types \cite{collin}.}
\label{models_table}
\begin {tabular}{|c|c|c|c|c|c|c|c|}
\hline
\multirow{2}{*}{Category} & Power & \multicolumn{3}{c}{ZIP - P} & \multicolumn{3}{|c|}{ZIP - Q}\\
\cline{3-8}
& factor & Z & I & P & Z & I & P\\
\hline\hline
GIL & 1 & 0 & 0 & 1 & 0 & 0 & 0 \\
CFL/LFL & 0.91$^{*}$ & -0.01 & 0.96 & 0.05 & 0.1 & -0.73 & -0.37 \\
LED LS& 0.137$^{*}$ & 0.23 & 0.85 & -0.08 & -1.05 & 0.04 & 0.01 \\
Resistive & 1 & 0 & 0 & 1 & 0 & 0 & 1 \\
$SMPS_{no PFC}$ & 0.994 & 0 & 0 & 1 & -3.63 & 9.88 & -7.25 \\
$SMPS_{pPFC}$ & 0.97 & 0 & 0 & 1 & 0.45 & -1.44 & 1.99 \\
$SMPS_{aPFC}$ & 1 & 0 & 0 & 1 & 0 & 0 & 0 \\
$RSIR_{QT}$ & 0.62 & 0.1 & 0.1 & 0.8 & 1.4 & -0.91 & 0.5 \\
$RSIR_{CT}$ & 0.62 & 0.63 & -1.2 & 1.57 & 1.4 & -0.91 & 0.5 \\
$CSCR_{CT}$ & 0.9 & 0.5 & -0.62 & 1.11 & 1.54 & -1.43 & 0.89 \\
\hline \multicolumn{8}{|l|}{where: $^{*}$ indicates a capacitive load. Note: the LED is lower power}\\
\multicolumn{8}{|l|}{variant intended for HIL spotlight replacement.}\\
\hline
\end{tabular}
\end{table}

\begin{figure}[h]
     \begin{center}
        \subfigure[Load types\cite{DECC}]{%
            \label{fig:load_typ}
             \includegraphics[trim=1.5cm 5cm 14cm 5cm, clip=true, width=0.47\textwidth]{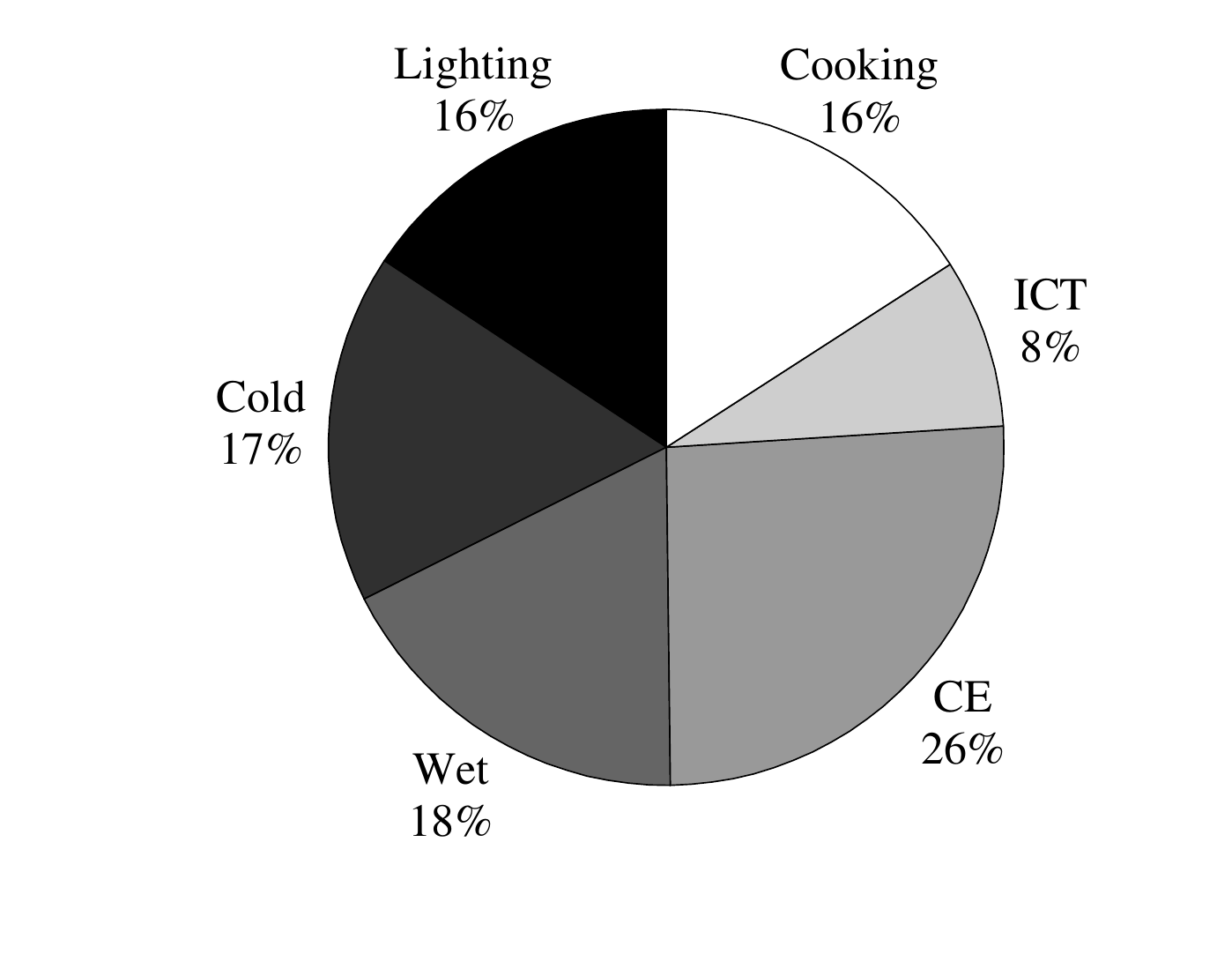} }
        \subfigure[Load categories]{%
            \label{fig:load_cat}
           \includegraphics[trim=14.2cm 5cm 2cm 5cm, clip=true, width=0.47\textwidth]{figures/load_cat_type.pdf} } \\
     \end{center}
     \caption{The energy consumption of each category and type of load in the UK households.}
     \label{fig:load_cat_typ}
\end{figure}

Figure \ref{fig:load_cat_typ} illustrates how the annual UK residential load sector consumption may be converted from load type into load categories. This offers a different representation of load statistics and can be directly obtained from the data presented in the main body text.

\subsection{Ownership statistics}
The final data set that is required to provide a comparitive study of the UK residential load sector is statistics of appliance ownership. This includes a high level of granularity, dividing each load type into a number of individual appliances. These are included in the developed pdf's to allow for direct implementation.

\section {Cold loads}
''Cold'' load covers all types of refrigerators and freezers. These appliances may be divided into four groups based on size and technical features \cite{LOT13}: group 1 includes small fridges (2 to 3 shelves) with or without a small freezer compartment ( Category 1 to 6); group 2 is defined as larger fridges ($>$3 shelves) with separate freezer section (Category 7 or 10). Stand-alone upright freezers are classified as group 3 (Category 8), while chest freezers are group 4 (Category 9). A review of the consumption, operation and ownership statistics is included at the end of this section in Table \ref{cold cycle}.

\subsection {Technical description}
These devices operate by circulating a refrigerant which absorbs heat from within the device and expends this via external heat exchange pipes. This cycle requires the refrigerant to change state, i.e. from gas to liquid and liquid to gas, and is achieved by a compressor and an expansion valve. The electrical demand is a result of the SPIM used to drive the compressor. This will cycle on and off as a result of thermostatic control to maintain the desired temperature, the typical cycle will last three quarters of an hour (with a duty ratio of approximately 0.33)\cite{synnpotential}.

As the compressors used in such devices do not require high starting or running torque, it is expected that 100\% of this load uses RSIR SPIMs. The motor load is a reciprocating compressor which behaves as a constant torque mechanical load \cite{motorInfo}. The power demand in operation stage (''on period'') varies between 25 to 252 W depending on the appliance and will not consume any power during in the ''off period'' \cite{LOT13,trends}. For all ''cold'' loads, the distribution of the rated power is considered to be normal, with further details in Table \ref{cold cycle}.

\subsection {Ownership statistics}
In the UK, every household will contain at least one ''cold'' load, defined as the primary device, and around a quarter of all households will have a secondary appliance \cite{LOT13}.

For the main/primary device, 20.7\% of fridges are within group 1 and the rest 79.3\% are from group 2. For the secondary refrigerator, a smaller appliance, both in terms of physical size and rated power, is mainly used since they consist of 80\% of the total number. For freezers, most households have an upright freezer (91.2\%) and 33\% possess a chest freezer.

\begin {table}[!ht]
\centering
\caption {Power ranges and ownership statistics for cold devices \cite{DECC,trends,LOT13}.}
\label{cold cycle}
\begin {tabular}{|c|c|c|c|c|c|}
\hline
\multirow{3}{*}{Load} & \multicolumn{2}{c|}{Device ownership (\%)} & \multicolumn{3}{c|}{Power Demand} \\
\cline{2-6}
& \multirow{2}{*}{Primary} & \multirow{2}{*}{Secondary} & On period & Off period & \multirow{2}{*}{Distribution} \\
& & & (W) & (W) & \\
\hline\hline
Group 1 & 20.7 & 80 & 36-45 & \multirow{4}{*}{0} & Normal: $\mu$ = 40, $\sigma$ = 1.5\\
Group 2 & 79.3 & 20 & 25-252 & & Normal: $\mu$ = 139, $\sigma$ = 23\\
\cline{2-3}
Group 3 & \multicolumn{2}{c|}{91.2} & 29-167 & & Normal: $\mu$ = 98, $\sigma$ = 23\\
Group 4 & \multicolumn{2}{c|}{33} & 28-186 & & Normal: $\mu$ = 107, $\sigma$ = 26\\
\hline \multicolumn{6}{|l|}{Note: the groups refer to definintions in Section 3 text.}\\
\hline
\end{tabular}
\end{table}

\section {Wet loads}
The ''wet'' load type consists of dishwashers, washing machines, tumble dryers and combined appliances, i.e. washer dryers. These appliances have specific operation cycles that vary in time duration, power consumption and model according to the device. A summary of all data is included at the end of this section in Table \ref{wet_stat}.

\subsection {Technical description}
\subsubsection {Washing machines}
The operation cycle of washing machines consists of two main stages stages. During the first stage, water is pumped into the drum and heated up to the required temperature using an ohmic heating element. In the next tage of operation, a sequence of drum rotations are performed with respect to the chosen cycle settings. The drum rotations will be alternated with pumping of fresh of water and flushing of water from the drum. The duration of the cycle depends on the selected washing programm and temperature and it may last from 15 min up to 3 hours \cite{synnpotential}. However, the typical cycle lasts for 75 minutes \cite{synnpotential,DsmoUK}. The load model variation of a typical cycle of a washing machine is depicted in Figure \ref{fig:wm_mod}.

The rated power of the heating element varies between 1.8 kW to 2.5 kW \cite{synnpotential} and is modelled as an ideal resistive load. The SPIM utilised to drive drum rotation is considered CSR$_{CT}$ due to the high running torque that is required. The centrifugal water pump will present a quadrative torque to the motor and is modelled as CSR$_{QT}$ \cite{motorInfo}. The power demand of the electronic control system of the device is very low and is assumed to include basic electronic components, and is, therefore, modelled as $SMPS_{no PFC}$.

\subsubsection {Dishwashers}
The dishwasher operation will contain three main stages. In the first stage, water is pumped into the device and heated to the required temperature. This is followed by several repititions, as set by the specific operating setting, of the cleaning process. In this stage, water is supplied to a spray fan element within the device. During the last stage, fresh water is drawn into the device and heated to complete the cleaning process. The duration of the cycle depends on the selected washing programm and temperature and it may last from 15 min up to 3 hours \cite{synnpotential}. However, the typical cycle lasts for 75 minutes \cite{synnpotential,DsmoUK}.

\begin{figure}[!h]
     \begin{center}
        \subfigure[Washing machine: power demand]{%
            \label{fig:wm}
            \includegraphics[trim=1.4cm 14cm 2cm 2.2cm, clip=true, width=0.5\textwidth]{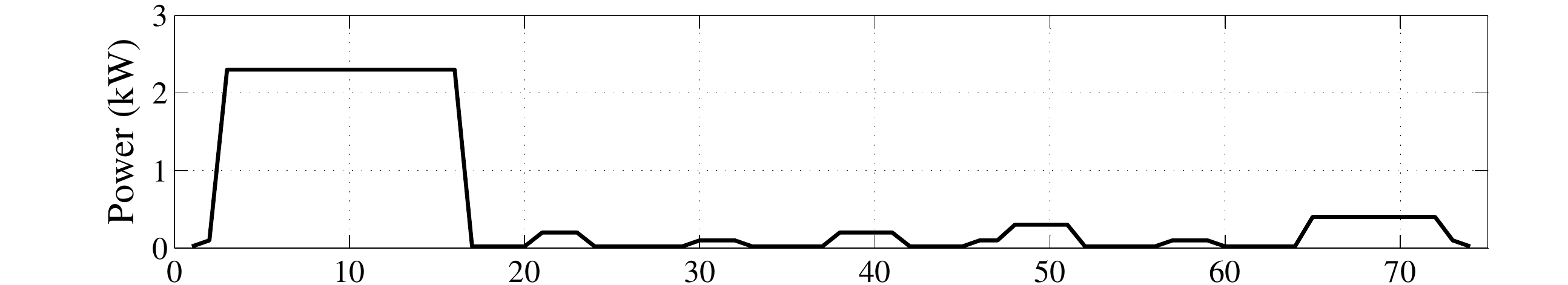} }
        \subfigure[Washing machine: model]{%
            \label{fig:wm_mod}
            \includegraphics[trim=1.6cm 14cm 2.5cm 1.5cm, clip=true, width=0.5\textwidth]{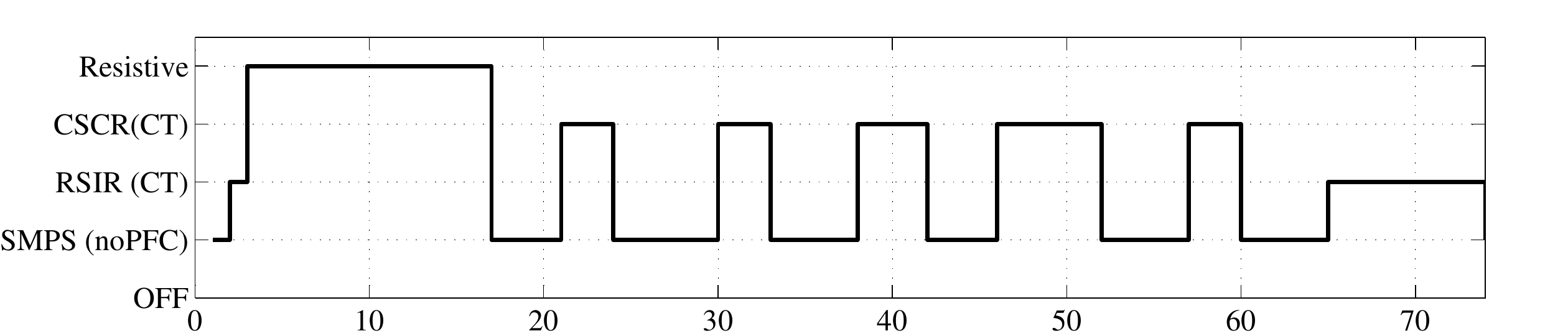} } \\ 

        \subfigure[Dishwasher: power demand]{%
            \label{fig:dw}
            \includegraphics[trim=1.4cm 7.8cm 2cm 8cm, clip=true, width=0.5\textwidth]{figures/power_var_wet_loads_2.pdf} }
        \subfigure[Dishwasher: model]{%
            \label{fig:dw_mod}
            \includegraphics[trim=1.6cm 7.8cm 2.2cm 8cm, clip=true, width=0.5\textwidth]{figures/model_wet_loads_2.pdf} } \\ 

        \subfigure[Tumble dryer: power demand]{%
            \label{fig:td}
            \includegraphics[trim=1.4cm 1.3cm 1.8cm 14cm, clip=true, width=0.5\textwidth]{figures/power_var_wet_loads_2.pdf} } 
        \subfigure[Tumble dryer: model]{%
            \label{fig:td_mod}
            \includegraphics[trim=1.6cm 1.3cm 2.2cm 13.8cm, clip=true, width=0.5\textwidth]{figures/model_wet_loads_2.pdf} } \\

        \subfigure[Washing dryer: power demand]{%
            \label{fig:wd}
            \includegraphics[trim=1.4cm 14cm 2cm 1.8cm, clip=true, width=0.5\textwidth]{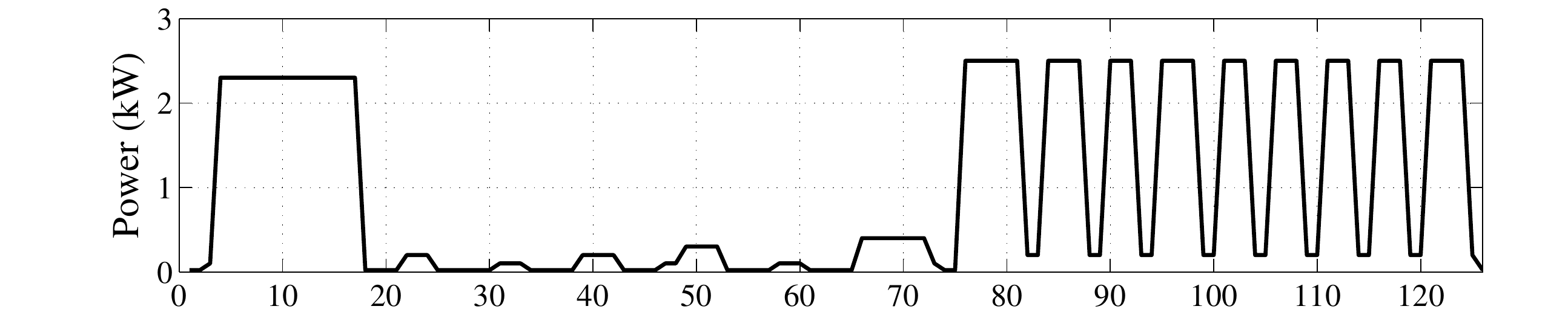} } 
        \subfigure[Washing dryer: model]{%
            \label{fig:wd_mod}
            \includegraphics[trim=1.4cm 14cm 2.5cm 2cm, clip=true, width=0.5\textwidth]{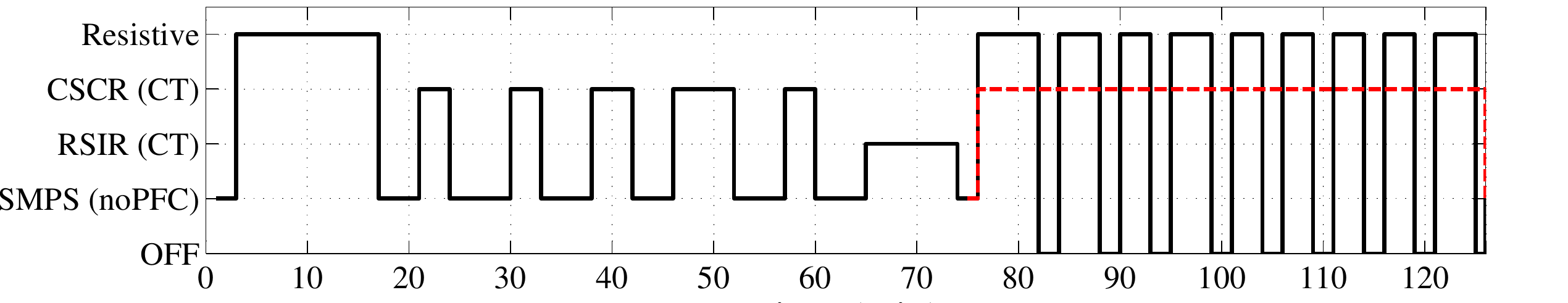} } \\
     \end{center}
     \caption{The operation cycles of the wet loads \cite{synnpotential} and the model variation during their operation.}
     \label{fig:subfigures_wet}
\end{figure}

As with the washing machine, the rated power of the heating element varies between 1.8 kW to 2.5 kW \cite{synnpotential} and the internal water pump is modelled as CSR$_{QT}$. Due to the lower running torque requirements it is assumed that the rotating cleaning fans does not include a run capacitor, i.e. it is modelled as $RSIR_{CT}$ load category. The power and load model variation of a typical dishwasher are depicted in Figures \ref{fig:dw} and \ref{fig:dw_mod} respectively.

\subsubsection {Tumble dryers}
The typical tumbe dryer contains only two main electrical components: a resistive heating element and a SPIM to drive drum rotation. The typical cycle consists of a periodic heating of the air within the appliance whilst the drum keeps rotating for the duration of the task \cite{synnpotential,REDDB}. The typical cycle lasts for 52 minutes \cite{synnpotential}. The rated power of the resistive heating element varies between 2 - 2.5~kW \cite{synnpotential}, while the motor for the drum rotation is modelled as $RSCR_{CT}$ due to the high running torque requirements (Figure \ref{fig:td_mod}). 

\subsubsection {Washer dryer}
A washer dryer is a washing machine that offers additional drying facitilies, similar to those performed in tumble-dryers. As such, it is assumed that the device operation will consist of the typical operating cycle of a washing machine followed by the typical operating cycle of the tumble-dryer.

\subsection {Ownership statistics}
The washing machine ownership in the UK is very high (around 93\%), while the penetration of dishwashers is more modest (about 35\%) \cite{DECC,LOT14}. As for the tumble dryers, about 69\% of UK households are equipped with one \cite{DECC}. The ownership statistics of washing machines and tumble dryers include the washer dryers. The ownership and the consumed energy per operation cycle of the wet loads are displayed in Table \ref{wet_stat}.

The ownership statistics and the consumed energy during the operation cycle for each cycle are shown in Table \ref{wet_stat}.

\begin {table}[!h]
\centering
\caption {Ownership statistics and consumed energy of operation cycle of wet loads \cite{DECC,LOT14}.}
\label{wet_stat}
\begin {tabular}{|c|c|c|}
\hline
\multirow{2}{*}{Load} & Ownership  & Power demand \\
& statistics (\%) & kWh/cycle \\
\hline\hline
Dishwasher & 35 & 1.1-1.8 \\
Washing machine & 93 & 0.6-1.5 \\
Tumble dryers & 69 & 2.4-4.2 \\
Washer dryers & 18 & 3-5.5 \\
\hline
\end{tabular}
\end{table}

\section {Consumer electronics}
The consumer electronics load type include a large number of different load types. TVs, VCR/DVD players, set-top boxes and all variants of power supplies units (PSU) are classified as consumer electronics. Outwith TVs, the majority of appliances are of low rated powers and can be simplified for modelling purposes. However, TV load requires more careful consideration.

\subsection {Technical description}
From an eletrical viewpoint, the required operation of all consumer electronic devices is identical. As they all require dc voltage to operate, all consumer electronic appliances require a SMPS to convert the ac supply voltage. Variations will be introducted based on the rated power of the device and the electrical circuits required to satisfy harmonic legislation, defined in cite{HarmLimits}. However, all of these devices will present a constant active power load on the power network.

\subsubsection{Low-power and supplementary devices}
All low-power and supplementary appliences, i.e.  set-top boxes, video players, game consoles, audio Hi-Fi and PSU, have rated power consumption less than or equal to the 75~W harmonic limit. Accordingly, they are modelled as PE no-PFC. Game consoles are similar to set-top boxes but with higher power consumption, between 19-197 W in operating mode depending on the console, therefore they are required to include some form of PFC circuit. Based on their high energy efficient operation, it is assumed that they are equipped with a-PFC.

\subsubsection{TVs}
In load use statistics, TVs are often referred to as primary and secondary. Primary TVs refer to the main use TV, while secondary TVs include smaller, lower rated power TVs with lower use frequency. Currently, there are four main variants of TV technology: cathode ray tube (CRT), liquid-crystal-display (LCD)/light-emitting diode (LED) display, plasma televisions and rear projectors (RP). The market share of the RP technology is negligible and is excluded from further analysis.

\begin{figure}[h]
\begin{center}
\includegraphics[trim=2cm 17cm 2cm 2.5cm, clip=true, width=0.8\textwidth]{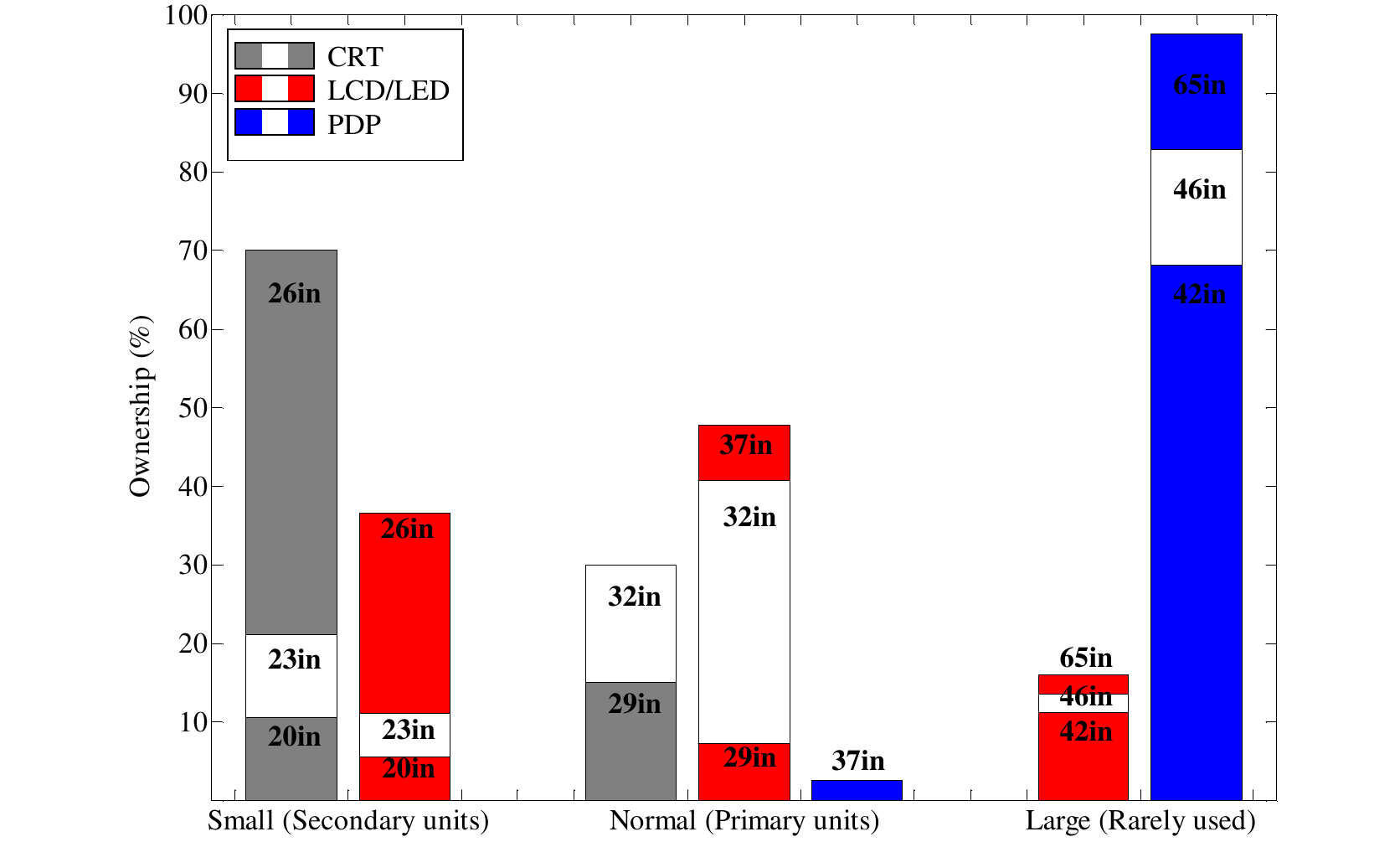}
\caption{TV ownership per screensize and technology\cite{DECC}.}
\label{fig:subfigures_TV}
\end{center}
\end{figure}

The rated power of the device will determine the electrical characteristics of the load, as a result of the implementation of harmonic legislation. All secondary appliances are considered as PE no-PFC, due to the lower rated power. For primary appliances, it is assumed that the modern appliances, i.e. LCD/LED, will utilise PE a-PFC technology, while older technologies, i.e. CRT and plasma, will include PE p-PFC circuits to satisfy harmonic legislation. The rated power of each of these appliances varies according to the size of screen and technology and can be selected from the rated power distribution, shown in Figure \ref{fig:subfigures_TV}. 

The distribution of rated power of the LCD/LED and Plasma TV technology is best represented by the Inverse Gaussian distribution, which is defined by two parameters: $\mu$ the mean value and shape factor $\lambda$ (\ref{eq:InverseGaussianDistribution}):

\begin{equation}
\label{eq:InverseGaussianDistribution}
f\left(x |  \mu, \lambda\right) = \sqrt{\frac{\lambda}{2\pi x^{3}}}e \left\{  -\frac{\lambda}{2\mu^{2}x}\left(x - \mu \right)^{2} \right\} \hspace{12pt}\text{for }\hspace{1pt} x > 0
\end{equation}

\subsection {Ownership statistics}
\subsubsection {TV's}
Almost all households are equipped with at least 1 TV, with the ownership to be about 97\% of households \cite{DECC}. A general rule of thum is that households do not have more than 1 appliance per occupant. In Figure \ref{fig:subfigures_TV}, it can be seen that secondary and most of the primary devices are LCD and, especially, CRT TV's which are the categories that concentrate the largest market share. Large screensize TV's consist of mostly plasma TV's and projectors and a few LCD. This distribution has a significant effect on the total power consumption of this load category.

\begin{figure}[h]
     \begin{center}
        \subfigure[LCD/LED TV power distribution.]{%
            \label{Fig:TVexampleLCD}
            \includegraphics[trim=0cm 1cm 1.5cm 0cm,clip=true, width=0.46\textwidth]{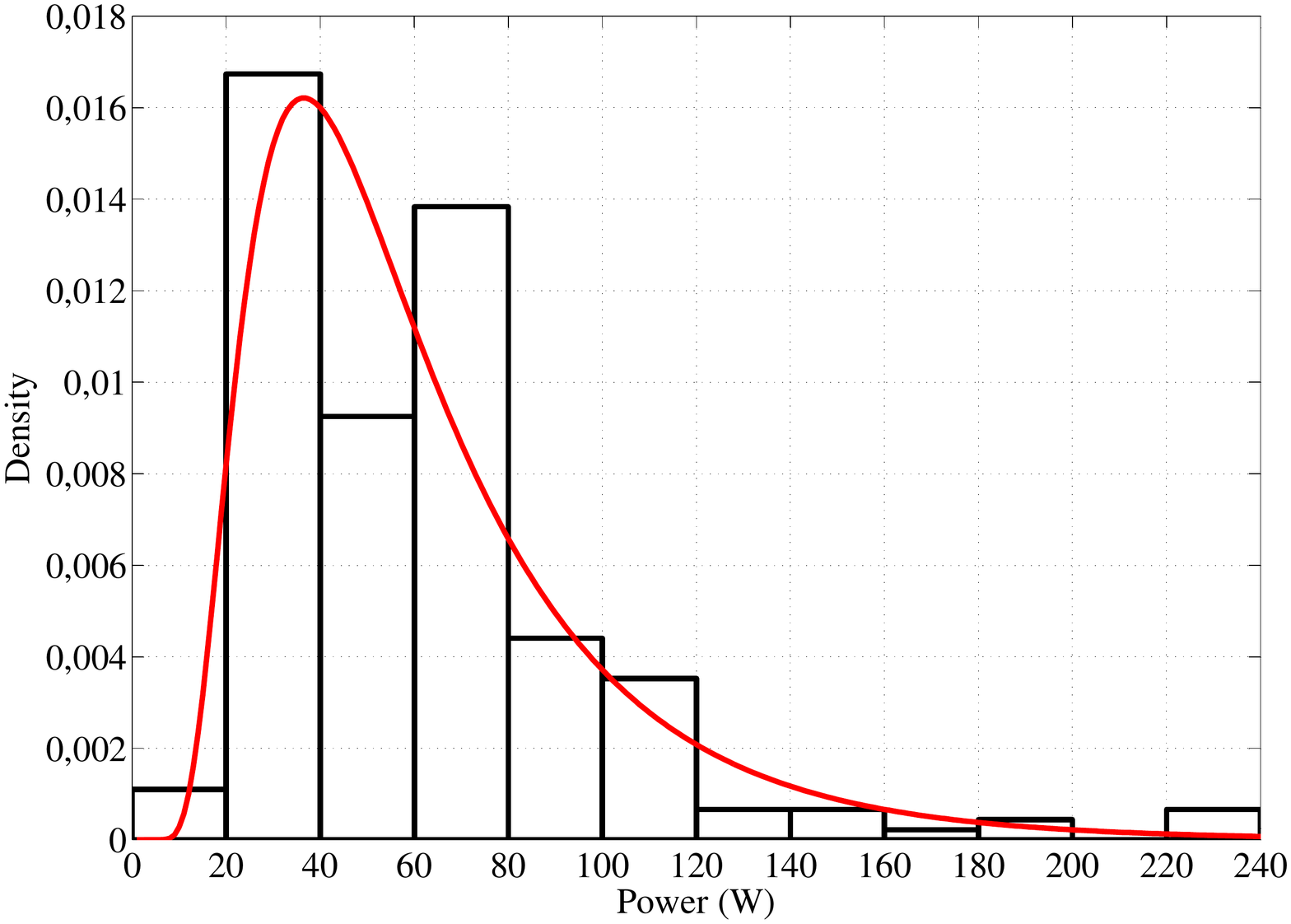} }
        \quad
        \subfigure[Plasma TV power distribution.]{%
            \label{Fig:TVexamplePlasma}
            \includegraphics[trim=0.2cm 0.85cm 0.1cm 0.24cm,clip=true, width=0.43\textwidth]{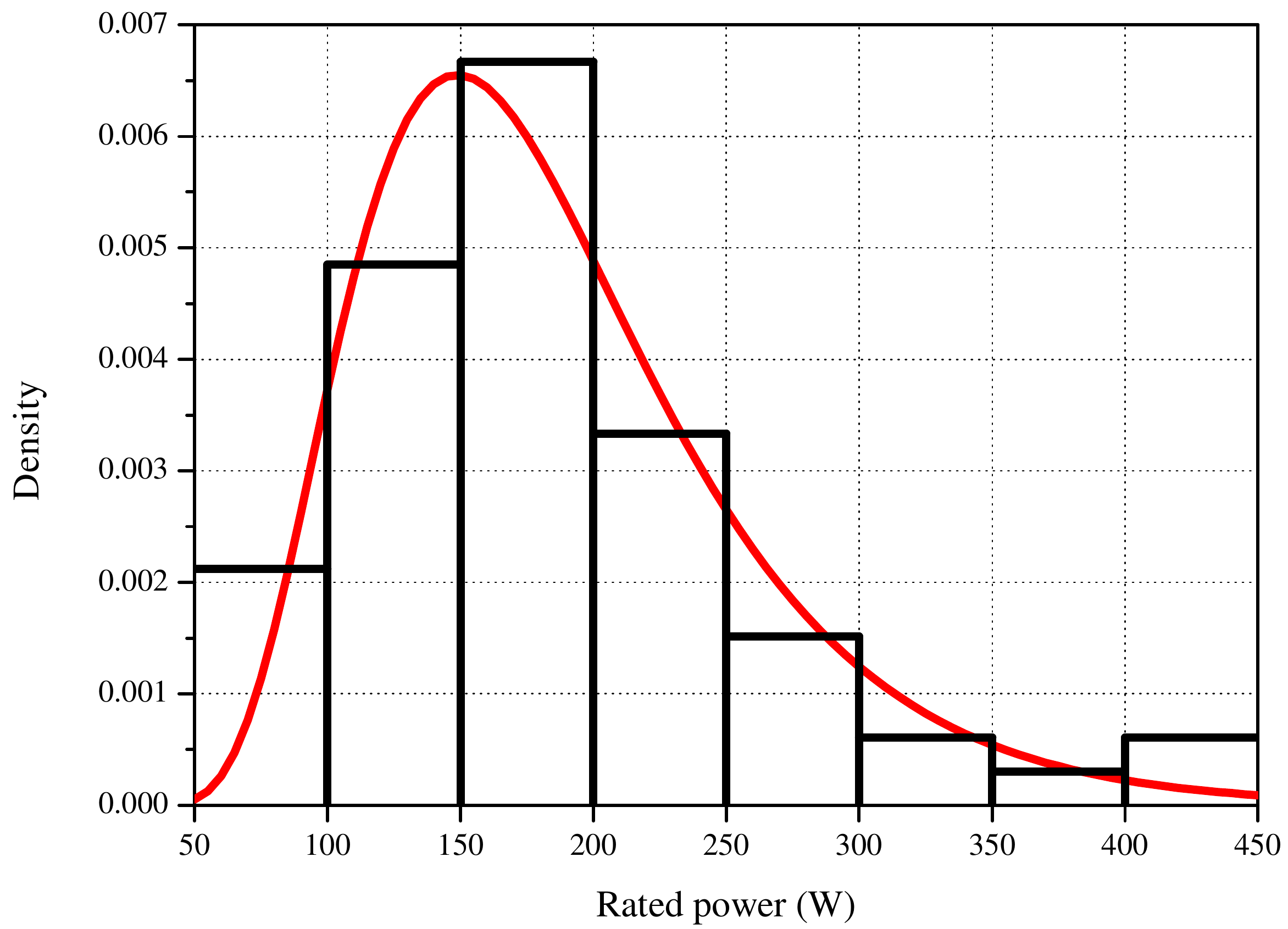} } \\ 
     \end{center}
     \caption{The distribution curve of rated power for LCD/LED and Plasma TV's.}
     \label{fig:tv_distrib}
\end{figure}

\begin {table}[h]
\centering
\caption {Statistics on TV ownership per technology and power range according to \cite{trends,Rmdeu}.}
\label{tv ownership}
\begin {tabular}{|c|c|c|c|c|}
\hline
\multirow{3}{*}{Device} & Device & \multicolumn{3}{c|}{Power demand}\\
\cline{3-5}
& ownership & In use& Stand-by&\multirow{2}{*}{Distribution}\\
& (\%) & (W) & (W)&\\
\hline\hline
CRT & 13.4 & 80-160 & 4-14 & Normal: $\mu$ = 60, $SD$ = 13.3\\
LCD/LED & 71.0 & 80-300 & 1-3 & Inverse Gaussian: $\mu$ = 60, $\lambda$  = 162\\
PDP & 15.1 & 60-130 & 2-4 & Inverse Gaussian: $\mu$ = 186, $\lambda$ = 1241\\
RP & 0.5 & 145-340 & 2-4 & - \\
\hline
\end{tabular}
\end{table}

\subsubsection {Supplementary and low power appliances}
As it is reported in \cite{3takeup}, a high proportion of UK households have set-top boxes and video-players, 93\% and 88\% respectively. The large penetration of set-top boxes can be attributed to the growing use of digitial television receivers. The ownership of audio appliances can be assumed to be a similar magnitude, around 90\%. Recent statistics indicate that approximately 44\% of UK households own a game console \cite{game_con}, with further details presented in Table \ref{gamecons ownership}.

\begin {table}[h]
\centering
\caption {Owneship statistics and rated power ranges for supplementary and low power appliances\cite{Rmdeu,stoch_model,EURECO}.}
\label{suppl_stat}
\begin {tabular}{|c|c|c|c|c|}
\hline
\multirow{3}{*}{Appliance} & Device &  \multicolumn{3}{c|}{Power demand} \\
\cline{3-5}
& ownership & In-use & Stand-by &\multirow{2}{*}{Distribution}\\
&(\%)&(W)&(W)&\\
\hline\hline
Set-top box & 93\% & 5-18 & 5-11 & Normal: $\mu$ = 12, $SD$ = 2.2\\
Video player & 88\% & 5-18 & 5-11 & Normal: $\mu$ = 12, $SD$ = 2.2\\
Audio appliances & 90\% & 18-30 & 5-6 & Normal: $\mu$ = 24, $SD$ = 2\\
Game consoles & 44\% & 13-197 & 2-4 & Uniform \\ 
\hline
\end{tabular}
\end{table}

\begin {table}[h]
\centering
\caption {Statistics on game consoles ownership per type and power consumption \cite{TVreport}.}
\label{gamecons ownership}
\begin {tabular}{|c|c|c|c|}
\hline
\multirow{3}{*}{Type}& Device  &\multicolumn{2}{c|}{Power demand}\\
\cline{3-4}&ownership&Mean Power& Stand-by Power\\
&(\%)&(W)&(W)\\
\hline\hline
PlayStation 3 & 27.9 & 23-197 & 1-3 \\
Xbox 360 & 29.5 & 67-185 & 2-3 \\
Nintendo Wii & 42.6 & 13-19 & 1-2 \\
\hline
\end{tabular}
\end{table}

\section {Information and Communication Technology}
Information and communication technology (ICT) load includes the following devices: PCs, monitors, laptops, printers and multi-function devices (MFDs). By annual power consumption, the main loads are PCs and laptops. All associated equipment with be operated in conjunction with one of these appliances. 

\subsection {Technical description}
As with the consumer electronics load type, the required operation of all ICT devices is identical, as all appliances will include a SMPS to convert the ac supply voltage to the required dc voltage.

\subsubsection {Desktop computers}
\begin{figure}[h]
\begin{center}
\includegraphics[trim=0cm 1cm 1.5cm 0cm,clip=true, width=0.7\textwidth]{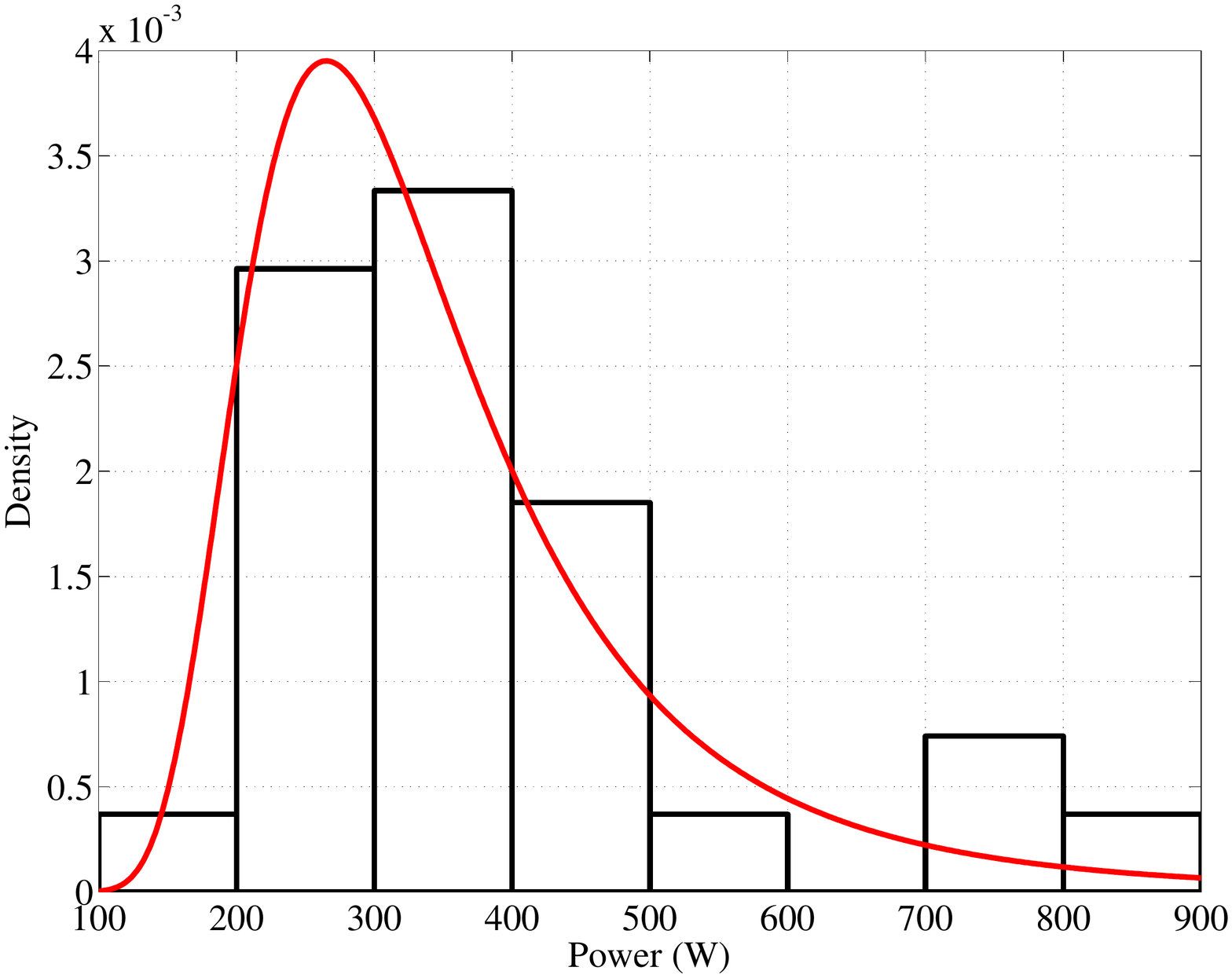}
\caption{Rated power distribution for desktops.}
\label{Fig:desktopDistr}
\end{center}
\end{figure}

Desktops are electronic devices that are assumed to have either passive PFC or active PFC depending on the circuit of the voltage supply. The active power consumption will vary in the range 20-60\% of the rated power during normal operating conditions, based on the specific operations being performed by the machine \cite{pc}. Although it is possible for the power demand to reach up to 100\% of the rated value, it is not expected to last for a significant period of time. According to manufacturers' datasheets, the rated power of the voltage supply varies between 100 to 900~W and the pdf can be described by the Generalised Extreme Value (GEV) distribution (Figure \ref{Fig:desktopDistr}). The GEV distribution is defined by three variables: the shape parameter $k$, the location parameter $\mu$ and the scale parameter $\sigma$ (\ref{GEVDistribution}):

\begin{align}
f\left(x\right) = \begin{cases} \frac{1}{\sigma} \left(-\left(1+kz\right)^{\frac{-1}{k}}\right)\left(1+kz\right)^{-1\frac{-1}{k}} & k \neq 0 \\
\frac{1}{\sigma}\exp\left(-z-\exp\left(-z\right)\right) & k = 0 \end{cases} 
\label{GEVDistribution}
\end{align} 

where 
\[z=\frac{x-\mu}{\sigma}  \hspace{12pt}\text{for } x > 0\]

\subsubsection {Monitors}
PC monitors’ electrical characteristics are similar to TVs but due to lower power demand, they are mainly modelled as $SMPS_{noPFC}$ for less than 75 W and $SMPS_{pPFC}$ for those that require more than 75 W. The required power depends on the age of the device, the technology utilised and the screen size.

\begin{figure}[!h]
\begin{center}
\includegraphics[trim=0cm 1cm 1.5cm 0cm,clip=true, width=0.7\textwidth]{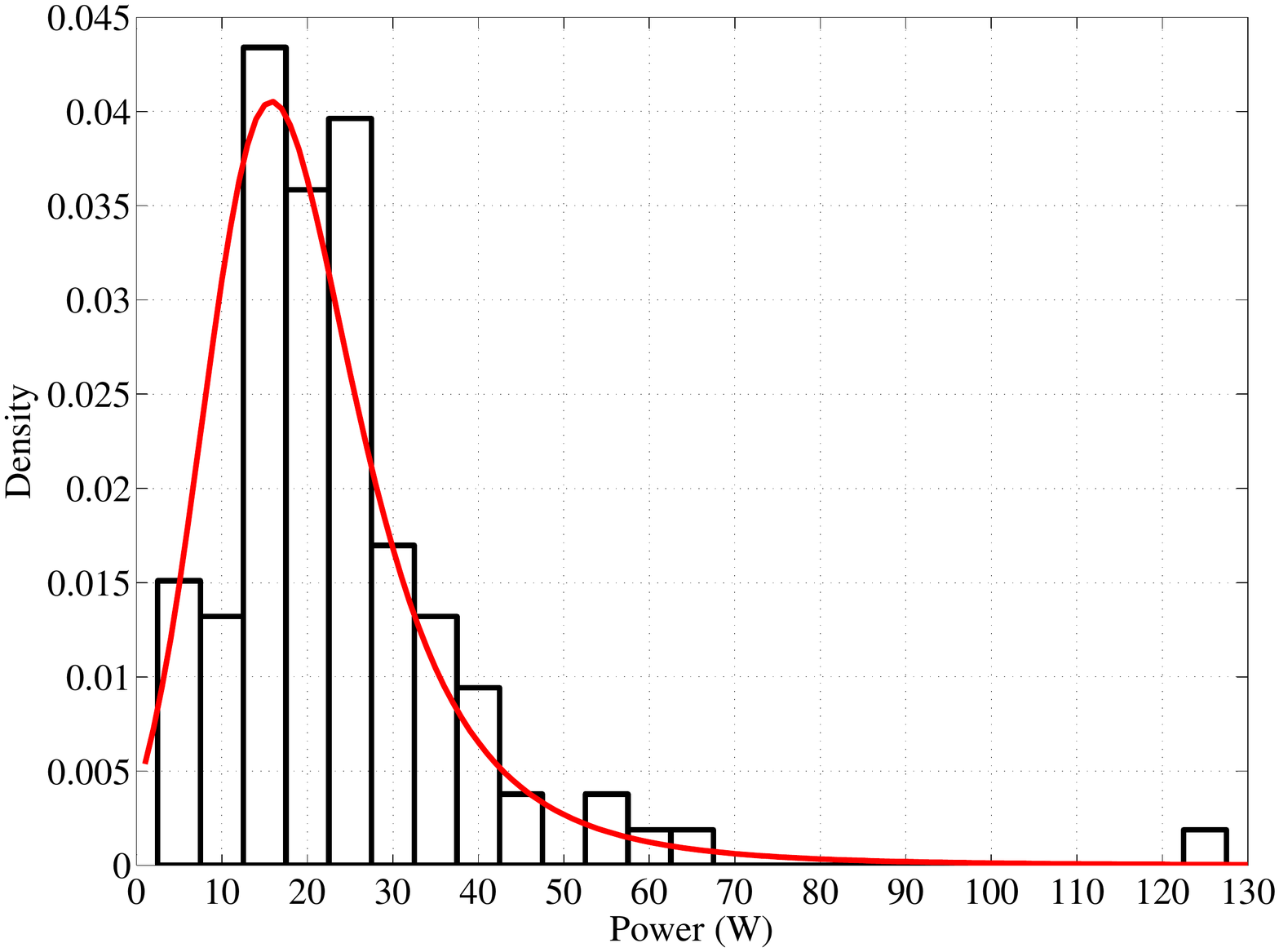}
\caption{LCD/LED PC monitors power distribution.}
\label{Fig:LCDMonitDistr}
\end{center}
\end{figure}

A typical CRT monitor will have active power rating in the range 60 - 85~W, drawing 2-5~W when in stand-by mode, based on their age. The power distribution of LCD/LED monitors can be described by the Log-Logistic distribution (eq.\ref{eq:LogDistribution}). This distribution is defined by three parameters: location parameter $\mu$ and scale parameters $\alpha$ and $\sigma$. The fitting to manufacturers' data is presented in Figure \ref{Fig:LCDMonitDistr}). LCD/LED monitor stand-by consumption is similar to CRT technology, 2-5~W \cite{Rmdeu,pc,ECbOaT}. 

\begin{equation}
\label{eq:LogDistribution}
f(x) = \frac{\alpha}{\sigma}\frac{(\frac{x-\mu}{\sigma})^{\alpha-1}}{(1+(\frac{x-\mu}{\sigma})^{\alpha})^2}  \hspace{12pt}\text{for }\hspace{1pt} x > 0
\end{equation}

\subsubsection {Laptops}
\begin{figure}[h]
\begin{center}
\includegraphics[trim=0cm 1cm 1.5cm 0cm, clip=true, width=0.75\linewidth]{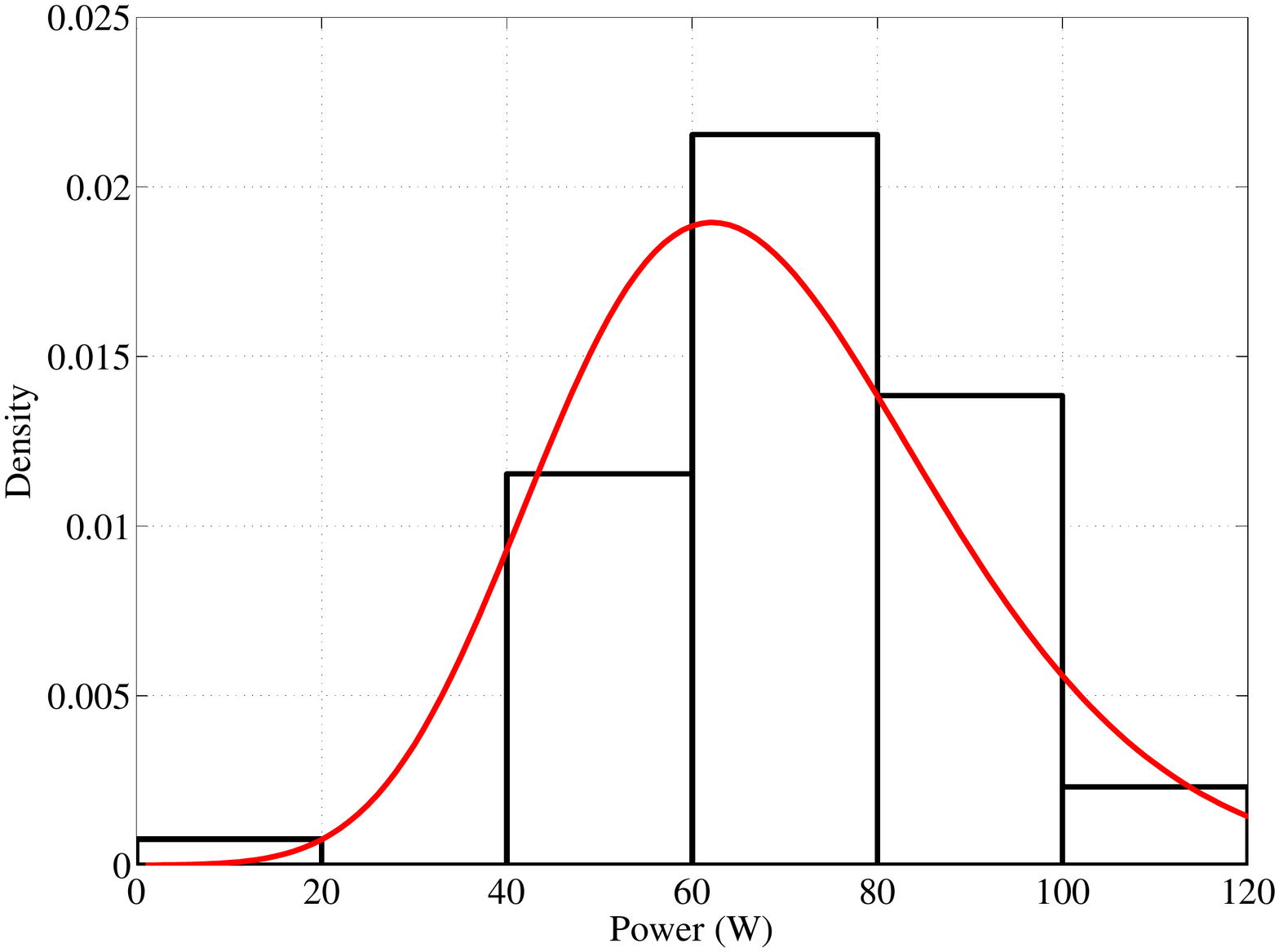}
\caption{Power distribution for laptops.}
\label{LaptopDistr}
\end{center}
\end{figure}

From a number of measurements performed on laptop battery chargers, with age ranging from one to six years old, it was found that the predominant technology was PE no-PFC ( 56\%), with PE p-PFC contibuting 19\% and PE a-PFC 25\% to this load type consumption. During operation, the power drawn by the laptop will vary between 40\% - 85\% of the rated power charger during normal operating conditions. The distribution of power drawn by the measured laptops and the laptops that are available in the market is shown in Fig.\ref{LaptopDistr}. This distribution can be represented by the GEV distribution.

\subsubsection {Office equipment}
This is classified as printers, scanners and MFD and there are several technological variations for each appliance type. However, from a load modelling viewpoint, this will only change the rated power of the device. From \cite{EURECO,ECbOaT} and manufacturers' datasheets, for ink-jet printers, the rated power will lie between 10-42 W, with 2-12W in stand-by mode. The power demand of laser printers is higher, typically between 315-440 W when active and 5-12 W in stand-by model. The active power demand of scanners is generally around 150~W, while MFDs will range from 15 - 500~W, depending on the size of the device.

It is assumed that all devices of rated power will utilised PE no-PFC technology, with passive-PFC being the main technology in devices of higher rated powers.

\subsubsection {Communication appliances}
Landline phones and routers are of low rated power, 7 - 30~W and are modelled as PE no-PFC \cite{pc}.

\subsection {Ownership statistics}
41\% of houses have a desktop and 78\% of them have laptops in UK, while about 30\% own both\cite{3takeup,DECC}. It is assumed that there is a maximum of one desktop in each household, with a maximum of one laptop per occupant. It is assumed that only 5\% of monitors currently in use are CRT technology, with the majority being LCD monitors.

Available statisics indicate that MFD are more widely used than standalone printers and scanners, with ownership approximatelt 57\% and 33\%, respectively. Router ownership is calculated in \cite{natMas} and confirmed by \cite{3takeup}, with approximately 84\% of households owning a router.

\begin {table}[h]
\centering
\caption {Owneship statistics and rated power ranges for ICT loads \cite{DECC,3takeup,pc,ECbOaT}.}
\label{ict_stat}
\begin {tabular}{|c|c|c|c|c|}
\hline
\multirow{3}{*}{Appliance} & Device &  \multicolumn{3}{c|}{Power demand}\\
\cline{3-5}
& ownership & In-use & Stand-by &\multirow{2}{*}{ Distribution} \\
&(\%)&(W)&(W)&\\
\hline\hline
Desktop& 41 & 50-250 & 5 & GEV: $k$ = 0.19, $\mu$ = 282 and $\sigma$  = 95\\
Laptops & 78 & 20-120 & 5 & GEV: $k$ = -0.15, $\mu$ = 58 and $\sigma$  = 20\\
CRT Monitors & 2.1 & 60-85 & 2-5& Normal: $\mu$= 72, $\sigma$ = 9\\
LCD Monitors & 39 & 10-140 & 2-5 & Log-Logistic: $\mu$ = 3.3, $\alpha$ = 29, $\sigma$ = 0.2\\
Printer/scanner & 32.5 & 10-440 & 2-15 & Inverse Gaussian: $\mu$ = 28, $\lambda$  = 148\\ 
MFD & 57 & 15-75 & 2-10 & Inverse Gaussian: $\mu$ = 25, $\lambda$  = 130\\ 
Routers & 84 & 12-40 & 2-3 & Normal: $\mu$ = 26, $\sigma$ = 4.7\\
Phones & 98 & 30-40 & 6-8 & Normal: $\mu$ = 35, $\sigma$ = 1.7\\ 
\hline
\end{tabular}
\end{table}

\section {Cooking}
Cooking loads can be divided into five general types: ovens, hobs, kettles, microwaves and small appliances. In the UK, the primary energy source for the major cooking loads, i.e. ovens and hobs, is gas, which is responsible for around 57\% of cooking energy. However, due to the high rated power of electric cooking appliances, there is still a considerable demand from electric cooking in the residential load sector. Recent statistics estimate that around 25\% of the daily power consumption comes from cooking \cite{BRE}.

\subsection {Technical description}
Cooking includes a large number of appliances that can be used for food preparation. The main electrical cooking devices are usually resistive loads which include ohmic heating element, and their rated power consumption depends on the type of appliance and their duty cycle ($D$). It is estimated that electrical hobs operate with $D=0.33$ for 0.75 min cycle, the values of $D=0.5$ for 5 min operation cycle for electric ovens are slight higher due to the higher temperatures required \cite{synnpotential}. The rated power of the resistive heating element between different appliances will varies in the range 2 to 3 kW.

The more complex operation of microwaves requires the rectification of the supply voltage and is modelled as PE p-PFC, while food processors are modelled as $CSCR_{CT}$ motors. Reports, e.g.
 \cite{DECC,trends,DsmoUK,EURECO,Decidomapp,ECCDA}, and manufactures' data give typical power range of the devices in market with a summary presented in Table \ref{cooking_stat}.

\subsection {Ownership statistics}
More than half of UK households (62\%) are equipped with electric ovens, while less than half (46\%) have electric hobs \cite{synnpotential,trends,DsmoUK}. Microwave ovens are more frequently found in UK dwellings, and the ownership is reported to be about 92\% \cite{DECC}. The penetration of smaller appliances, i.e. kettle, a toaster or a food processor, is considered to include all UK households \cite{DECC}. Table \ref{cooking_stat} presents the ownership statistics and the rated power ranges, given as uniform distribution.

\begin {table}[h]
\centering
\caption {Cooking appliance ownership  \cite{DECC,synnpotential,trends,DsmoUK,EURECO,Decidomapp,ECCDA}.}
\label{cooking_stat}
\begin {tabular}{|c|c|c|c|}
\hline
\multirow{3}{*}{Device} & Device &  \multicolumn{2}{c|}{Power demand}\\
\cline{3-4}&ownership&Rated power &\multirow{2}{*}{Distribution}\\
&(\%)  & (kW) & \\
\hline\hline
Electric oven & 62 & 2-3 & Normal: $\mu$ = 2500, $\sigma$ = 155\\
Electric hob & 46 & 2-3 & Normal: $\mu$ = 2500, $\sigma$ = 155\\
Microwave Oven & 92 & 0.6-1.15 & Normal: $\mu$ = 862, $\sigma$ = 97\\
Kettle & 98 & 2-3 & Normal: $\mu$ = 2500, $\sigma$ = 167\\
Toaster & 95 & 0.8-1 & Normal: $\mu$ = 900, $\sigma$ = 33\\
Food processor & 95 & 0.15-0.33 & Normal: $\mu$ = 240,$\sigma$ = 30\\
\hline
\end{tabular}
\end{table}

\section {Shower}
Similar to the cooking activity, it is possible to use electrical showers, which are essentially instantaneous water heaters (see Section  11), or to draw hot water from a gas boiler system.

\subsection {Technical description}
The water is heated up instantly by a heating element with high rated power that may vary between 4 to 9 kW \cite{trends}. It is assumed that the power consumption remains constant during the activity since there is not any tank to store the hot water and new, cold water is constantly drawn into the appliance. The electric shower is modelled as a constant resistance load due to the use of resistive heating element.

\subsection {Ownership Statistics}
It is estimated that about 49\% of UK households use electricity to heat the water and 90-92\% of them are equipped with electric showers \cite{synnpotential,LOT2}.

\section {Miscellaneous loads}
This load type covers all small devices that either have low power consumption or have low frequency of use. In this report, this includes: vacuum cleaners, irons and hairdryers. These appliances usually have high rated power but they are used less frequently and for short periods of time.
\subsection {Technical description}
Vacuum cleaners consist of a $CSCR_{CT}$ motor because of the high running torque that is required. An iron is mainly a large resistive heating element while a hairdryer model is a combination of both. Hairdryers consist of a high running torque motor and a resistive element to heat the air. A summary of manufacturers' data is presented in Table \ref{misc}.

\subsection {Ownership Statistics}
The penetration of vacuum-cleaner in market is 93.7\% \cite{vacuum_cl}. Similarly, steam irons and hairdryers are assumed to be in the majority of households.

\begin {table}[h]
\centering
\caption {Power demand and ownership statistics for miscellaneous appliances.}
\label{misc}
\begin {tabular}{|c|c|c|c|}
\hline 
\multirow{3}{*}{Device} & Device &\multicolumn{2}{c|}{Power demand}\\
\cline{3-4}&ownership & In-use power & \multirow{2}{*}{Distribution} \\
& (\%) & (kW) & \\
\hline\hline
Vacuum cleaner & 93.7 & 1.5-2.5 & Normal: $\mu$ = 2, $\sigma$ = 0.2\\
Iron & 95 & 2-2.8 & Normal: $\mu$ = 2.4, $\sigma$ = 0.1\\
Hairdryer & 95 & 1.8-2.2 & Normal: $\mu$ = 2, $\sigma$ = 0.1\\
\hline
\end{tabular}
\end{table}

\section {Lighting}
\subsection {Technical description}
GILs have an electrical filament in a glass bulb filled with inert gas that is heated to a high temperature until it glows. Halogen lamps are very similar to the GIL's. They are also made of a filament inside a glass bulb that contains halogen gas. The chemical reaction between the filament and the gas allows for higher luminous efficacy for the same amount of power in comparison with GIL. Therefore, the electrical characteristics will be similar. The fluorescent lamp is a gas-discharge lamp that uses electricity to excite mercury vapor. To iniate and control this process, a sophisticated ballast circuit is required. There are two types of fluorescent lamps according to the shape: linear (LFL) and compact (CFL). LED LS are the most technologically advanced type of lamp which consists of light-emitting diodes and an internal (or external) rectifier to supply them with direct current (DC). 

Due to technological differences, each type of lamp is modelled requries a unique load model (see Table \ref{models_table}). Further details are provided in \cite{collin,Djokic}.

\subsection {Ownership Statistics}

Lighting is assumed to consume approximately 20\% of the total residential power consumption \cite{DECC,lighting}. In 2010, CFL, halogen and GIL lamps are the three dominant types of lamps in the UK market, with market shares of each type given in Table \ref{table_lighting} \cite{DECC,lighting}. Given that GIL share is being reduced gradually over time, the share of CFL and halogen lamps will increase further in the future. 

\begin {table}[h]
\centering
\caption {Percentage and power model of lamp types in stock \cite{collin,DECC,Djokic,lighting}.}
\label{table_lighting}
\begin {tabular}{|c|c|c|}
\hline
\multirow{2}{*}{Type} & Share & Power \\
& (\%) & Range (W) \\
\hline\hline
GIL 40W & 16.2 & 40 \\
GIL 60W & 16 & 60 \\
GIL 100W & 2.8 & 100 \\
Halogen & 27.4 & 5-500 \\
LFL & 2.7 & 80-150 \\
CFL & 34.3 & 8-23 \\
LED & 0.6 & 4-13 \\
\hline
\end{tabular}
\end{table}

\section {Water heating}
For domestic water heating, gas and electricity are the primary energy source for approximately all UK households. The electrical water heating market is evenly divided into instantaneous water heating and water storage heating \cite{LOT2}.

\subsection {Technical description}
In instantaneous water heating, a small tank (smaller than 30 litres) is used, where the water is heated and used as it is required. It is similar in operation to an electrical shower, apart from the tank. In this appliance, whenever hot water is consumed, it is withdrawn from the tank which is instantly heated. 

A water storage heating system consists of a larger tank for the hot water. The large amount of energy required to maintain this large volume of water at a maximum set temperature makes it a significant electrical load. Consequently, it is used mainly by houses that have Economy 7 tariff and heat up the water during night so as to consume it during the day \cite{water_stat}. During the day, if the temperature of the water falls under a defined lower set temperature, the heater operates to raise the temperature to the desired delivered temperature \cite{water_stat}.

The heater is modelled as a resistive element while the centrifugal pumps used for circulating the water, are modelled as $CSCR_{QT}$ as high running constant torque is needed to elevate the water up to the required height. From an exhaustive review of manufacturers's data, the power demand of pumps varies depending on the size, with pdf described by GEV distribution (eq. \ref{GEVDistribution}) in the range of 20-245 W. The rated power of instantaneous water heaters that are currently available varies between 2 to 6 kW. The typical rated power of currently available heating element in water storage heaters (as a function of tank volume) and ownership statistics are presented in Table \ref{heaters ratings} \cite{LOT22}.

\subsection {Ownership statistics}
In the UK, 48\% of houses use electricity to heat water for daily domestic use. Excluding the showers (90-92\%), it is estimated that the market is split in half between instantaneous ($<$30 litres tank) and storage ($>$30 litres tank) water heating systems \cite{LOT2}. 

\begin {table}[!h]
\centering
\caption {Volume, rated power of available water storage heaters and ownership statistics.}
\label{heaters ratings}
\begin {tabular}{|c|c|c|}
\hline
Volume & Device & Rated \\
(lt) & Ownership (\%) & power (kW)\\
\hline\hline
80 & 12 & 2 \\
100 & 35 & 3 \\
150 & 15 & 4 \\
200 & 34 & 5 \\
400  & 4 & 6 \\
\hline
\end{tabular}
\end{table}

\section {Space heating}
The principal energy source for space heating in the UK is gas. However, some households will use resistive heating systems in conjunction with off-peak electricity tariffs. A more common form of electrical space heating is in the form of electric storage heaters and there is a small share of smaller/portable instantaneous electrical heaters.

\subsection {Technical description}
The rated power of electric space heating ranges between 5 to 7 kW with uniform distribution \cite{synnpotential} and modelled as a resistive element. The storage heaters need to be charged and then they heat the room by using the stored energy. They require a lot of time to chage and, consequently, they are a large resistive load which is mainly installed in houses that have Economy 7 tariff and are charged during night.

\subsection {Ownership Statistics}
As stated previously, gas is the dominant energy source for space heating in the UK residential load sector. However, there is a small proportion of houses, approximately 7\%, that are equipped with electrical space heating. 95\% out of them have installed storage heating systems and only 5\% is using instantaneous electrical heaters \cite{DECC}.

\section{Summary}
This paper compiles and summarises the statistical data of the most widely used electrical loads in the UK residential load sector. This review includes data from publically available references, an exhaustive review of manufacturers' data and device measurements. The data age varies between 1 - 5 years, which is considered acceptable for drawing conclusions on the existing technology of the devices and ownership statistics.

In this paper, the UK residential load sector was divided into the main individual loads. For every load, probability distribution functions for the active power demand were developed from the manufacturers data and ownership statistics presented. The approach applied in this paper offers additional data, which is not included in existing load consumption statistics, on the electrical characteristics of the loads. This holistic approach includes consideration of the technical operation of each appliance to identify the specific components which perform the electrical processes requried by the load. This is particularly important for the application of load use statistics in power system analysis.

\bibliographystyle{IEEEtran}
\bibliography{Tsagarakis_et_al_Domestic_Appliances}

\end{document}